\documentclass[11pt]{amsart}
\usepackage[utf8]{inputenc}
\usepackage{upgreek}
\usepackage{amsmath, amsfonts, amsthm}
\usepackage{upgreek}
\usepackage{graphicx}
\usepackage{subfigure}
\usepackage{todonotes}
\usepackage{caption} 
\usepackage[symbol]{footmisc}

\title{Optimizing Bone Scaffold Porosity Distributions}
\author[PSP Poh]{Patrina S.P. Poh}
\address{Patrina S.P. Poh\footnote{PSP Poh and P Dondl contributed equally to this work. \label{foot:authors}}; Department of Experimental Trauma Surgery; Klinikum Rechts der Isar; Technische Universit{\"a}t M{\"u}nchen; Munich; Germany and Julius Wolff Institute for Biomechanics and Musculoskeletal Regeneration; Charit{\'e} -- Univerist{\"a}tsmedizin Berlin; Berlin; Germany}
\author[D Valainis]{Dvina Valainis}
\address{Dvina Valainis; Department of Experimental Trauma Surgery; Klinikum Rechts der Isar; Technische Universit{\"a}t M{\"u}nchen; Munich; Germany}
\author[K Bhattacharya]{Kaushik Bhattacharya}
\address{Kaushik Bhattacharya; Division of Engineering and Applied Sciences; California Institute of Technology; Pasadena, CA; USA}
\author[M van Griensven]{Martijn van Griensven}
\address{Martijn van Griensven; Department of Experimental Trauma Surgery; Klinikum Rechts der Isar; Technische Universit{\"a}t M{\"u}nchen; Munich; Germany}
\author[P Dondl]{Patrick Dondl}
\address{Patrick Dondl\textsuperscript{\ref{foot:authors}} (corresponding author); Abteilung f{\"u}r Angewandte Mathematik; Albert-Ludwigs-Universit{\"a}t Freiburg; Freiburg; Germany}
\email{patrick.dondl@mathematik.uni-freiburg.de}
\date{\today}

\begin{document}
\begin{abstract}
	We consider a simple one-dimensional time-dependent mod\-el for bone regeneration in the presence of a bio-resorbable polymer scaffold. Within the framework of the model, we optimize the effective mechanical stiffness of the polymer scaffold together with the regenerated bone matrix. The result of the optimization procedure is a scaffold porosity distribution which maximizes the stiffness of the scaffold-bone system over the regeneration time, such that the propensity for mechanical failure is reduced.
\end{abstract}
\maketitle

\section{Introduction} \label{sec:intro}
The regeneration and restoration of skeletal functions of critical-sized bone defects are very challenging. Presently, regenerative therapy using biodegradable scaffolds have shown promising results in vivo~\cite{Berner:2013,Cipitria:2013,Reichert:2011,Sawyer:2009,Rai:2007} and in clinical cases~\cite{Ip:2007,Hoda:2016,Teo:2015fu,Schuckert:2008hg}. During the bone regeneration processes, scaffolds act as a temporary supporting structure to \textit{(a)} ensure that the defect/regeneration space is suitable for bone tissue growth, maturation and remodeling; \textit{(b)} provide mechanical functionality for proper transfer of loads acting on scaffolds to the adjacent host tissues while tissue regenerates; and \textit{(c)} facilitate in-growth of tissue and vasculature to accelerate tissue regeneration. Hence, the scaffolds' architecture plays a crucial role during bone regeneration processes. Fundamentally, in the design of bone scaffolds, there are several functional characteristics to be considered, such as the porosity (or biomaterial volume fraction), pore size and pore shape, as these will affect the scaffolds’ permeability/diffusivity, degradation rate and elastic modulus, and in turn the biological processes necessary for regeneration~\cite{Poh:2016cu}.
 
Due to the intricate relationships between scaffold geometries, mechanical properties, biomaterials and biological processes, research in bone tissue engineering has been dominated by trial-and-error approach---whereby an existing design is modified based on the experimental outcomes. This approach usually requires costly protocols and time-consuming experiments. Over the years, due to rapid development of computer-aided engineering (CAE) tools, topology optimization techniques have shown their potential as a powerful tool for the design of scaffolds’ architecture. Topology optimization is a numerical process to iteratively distribute a given amount of material within a given design space under specific constraints such that the final structure meets specific design targets~\cite{Allaire:2012ia,Bendsoe:2013ek}. This technique has been applied to design scaffolds to topologically achieve the optimized architecture to match the desired porosity, elastic modulus and fluid permeability~\cite{Dias:2014, Coelho:2015, Lin:2004, Guest:2006, Challis:2012, Kang:2010}.
 
One limitation common to most CAE models for topology optimization of bone scaffolds architectures is that the time-dependence of the regeneration process is not taken into account. Bone regeneration using scaffolds is a complex phenomenon which involves different biophysical processes and scaffold/tissue interactions via their elastic moduli over time. To this end, various multi-scale models have been proposed and developed simulating bone regeneration processes in response to various scaffolds properties (i.e., porosity, permeability, elastic modulus), taking into account the time-evolution of the microstructure due to bone growth and scaffold resorption~\cite{Sanz-Herrera:2008}. Hollister et al.~\cite{Hollister:2002} proposed a topology optimization approach that can design a scaffold microstructure in order to meet the resulting conflicting design requirements. 

However, implementation of CAE into routine additive manufacturing (AM) workflows for optimization of scaffolds for regenerative medicine is impeded by the high computational cost. Here, we propose a simple, one-dimensional model for the optimization of bone scaffold architectures based on homogenized quantities. The main advantage of this model will be its efficiency for numerical computation. We can therefore use it as a first test case for a scaffold shape-optimization procedure, in particular with respect to finding suitable objective functions to optimize. Our approach is thus related to the first step in the ``Shape Optimization by the Homogenization Method''~\cite{Allaire:2012ia} of optimizing within a relaxed problem for averaged quantities.

The remainder of this article is organized as follows. In section~\ref{sec:model} we introduce our simplified bone regeneration model and show some comparison to in-vivo studies. Section~\ref{sec:opt} describes the numerical optimization routine. The results of our numerical experiments are presented in section~\ref{sec:num}. Finally in section~\ref{sec:conc} we discuss conclusions and give an outlook to future work.

\section{Development of a scaffold-mediated bone regeneration model}\label{sec:model}
In bone tissue engineering, the process of bone regeneration is commonly mediated by osteoconductive scaffolds that are often combined with growth factors and/or cells (osteoinductive). However, the exact mechanism by which scaffold-mediated bone regeneration occurs is yet unclear. The process of bone regeneration is a complex and continuous process, but well orchestrated, starting from the formation of hematoma accompanied by the infiltration of inflammatory cells. Following the events of inflammation, bone regeneration can occur through either endochondral (bone formation through an intermediate cartilage phase) or intramembranous ossification (formation of new bone directly - usually occuring adjacent to existing bone). Finally, the regenerating bone enters the remodeling stage, where newly formed bone continues to remodel itself until a mechanically strong and highly organized bone structure is restored. The specific mechanism of bone regeneration is determined by the immediate biomechanical and chemical environment where the cell resides. For example, studies of bone regeneration mediated by scaffolds made from a composite of polycaprolactone (PCL, a slow degrading synthetic thermoplastic) and $\upbeta$-tricalcium phosphate ($\upbeta$-TCP) illustrated that porous scaffolds support formation of a structured fibrous tissue across the defect, which acts as the supporting network guiding the mineralization process and bone in-growth through the depth of the defect~\cite{Cipitria:2012}. More recently, a follow-up study revealed a non-standard form of mineralization of extracellular matrix within a porous scaffold, whereby extracellular matrix (ECM) mineralization occurs without collagen remodeling and without an intermediate cartilage ossification phase, extending from the proximal and distal end of the osteotomy site parallel to the tibia axis~\cite{Paris:2017}. In both studies~\cite{Cipitria:2012, Paris:2017}, it was noted that in the absence of exogenous growth factors or cells, no clinically relevant bridging of the defect was achieved.

In this study, the primary intention of the proposed scaffold-mediated bone regeneration model is to provide a basis for the scaffold architecture optimization algorithm proposed in section 3.  Hence, we propose a simple mathematical model capturing specific events occurring during the process of scaffold-mediated bone regeneration. However, the model does take into consideration the time-aspect of tissue regeneration, considering both bone growth and scaffold degradation and the interaction of the two. The time evolution model tracks the time dependent quantities that are relevant for the mechanical integrity of the scaffold. These quantities are the molecular weight of the scaffold material (which diminishes over time due to degradation) and the amount of bone regenerated. The bone regeneration, however, depends on the local microenvironment. This includes \textit{(a)} the presence of endogenous angiogenic and osteoinductive factors (e.g., growth factors/cytokines), which  are excreted into, diffuse through, and decay in the extracellular matrix in the interstitial space creating a local gradient through the regenerating tissue; and \textit{(b)} the mechanical strain stimulus transmitted through the structure due to external mechanical loading as it is well established that mechanical stimuli play a major role in bone regeneration~\cite{Ghiasi:2017aa}. 

 A coupled system of evolution equations for these quantities, together with boundary conditions, has been established and constants in the model (e.g., the degradation rate of the scaffold material) have been deduced from experimental observations. We note that the model does not resolve these quantities on a fine ($\upmu m$) scale, but uses coarse-grained values. The spatial domain of computation is the domain occupied by the scaffold. We simplify this domain to a one-dimensional object by only considering one of the main stress axis under physiological mechanical loading.

Concretely, we solve the following system of differential equations (in space on the domain $(0,L)$ and time for $t\in (0,T]$) for the scaffold volume fraction $\rho=1-\theta$ with $\theta$ being the scaffold porosity, the relative molecular weight of the scaffold material $\sigma$ (normalized to be equal to unity for a new scaffold), the relative bone density $b$, the density of active biological molecules (osteoinductive factors) $a$, and the mechanical displacement $u$:
\begin{align*}
\rho_t &= 0 & \parbox{20em}{(occupied space does not change in time)}\\
\sigma_t &= -k_1 \sigma & \parbox{20em}{(exponential loss of molecular weight)} \\
a_t &= (D(\rho)a_x)_x + k_2|u_x| b - k_3a & \parbox{20em}{(diffusion, generation, and decay of \\active molecules)} \\
0 &= (\mathbb{C}(\rho,\sigma,b)u_x)_x & \parbox{20em}{(mechanical equilibrium)}\\
b_t &= k_4 a|u_x|\left(1-\frac{b}{1-\rho}\right) & \parbox{20em}{(bone growth proportional to $a$ and\\ mechanical strain but constrained by\\ free volume)} 
\end{align*}
In this system, $k_1$, $k_2$, $k_3$ and $k_4$ are constant parameters, $D$, and $\mathbb{C}$ are functional relationships, all to be determined by experiment.

This model assumes PCL as the scaffold material. In vivo, PCL showes a two-stage degradation pattern, predominantly by bulk erosion. The first stage involves a decrease in molecular weight (Mw) without volume loss. The second stage of degradation begins when the Mw drops below 8000 Da, at which the material becomes brittle and loss of volume occurs~\cite{Sun:2006}, but this happens past the time scale considered here. Therefore, we consider no loss of volume fraction of the scaffold material to occur over time.

The initial and boundary conditions for the respective variables are given by
\begin{align*}
\rho(x,0) &= \rho_0(x) & \text{for all $x\in (0,L)$}\\
\sigma(x,0) &= 1 & \text{for all $x\in (0,L)$} \\
a(x,0) &= 0 & \text{for all $x\in (0,L)$} \\
a(0,t)=a(L,t) &= 1 & \text{for all $t\in [0,T]$}\\
u(0,t)=0, u(L,t) &= \gamma L & \text{for all $t\in [0,T]$}\\
b(x,0) &= 0 & \text{for all $x\in (0,L)$},
\end{align*}
i.e., in the beginning, the scaffold is fully intact, and no regenerated bone is present. The bone adjacent to the scaffold, however, produces the active molecules such that saturation is achieved there. The value for $u(L,t)$ was chosen such that the scaffold is subject to a hard-loaded engineering strain $\gamma$\footnote{We note that the magnitude of the displacement boundary condition is arbitrary and could be normalized to unity by changing the parameter $k_4$. Furthermore, the displacement $u$ is non-dimensionalized.}.

The resulting evolution equations are solved using established numerical methods, in our case, an implicit first-order Euler scheme. Due to the coarse graining of the relevant variables in the model, such a computation can be done in a very short time.

\begin{figure}
\centering 
\subfigure[Relative bone density at $t$=12 months with initial condition $\rho_0=0.1$ and $\rho_0=0.7$, respectively.]{\label{fig:b_unopt}\includegraphics[width=0.45\textwidth]{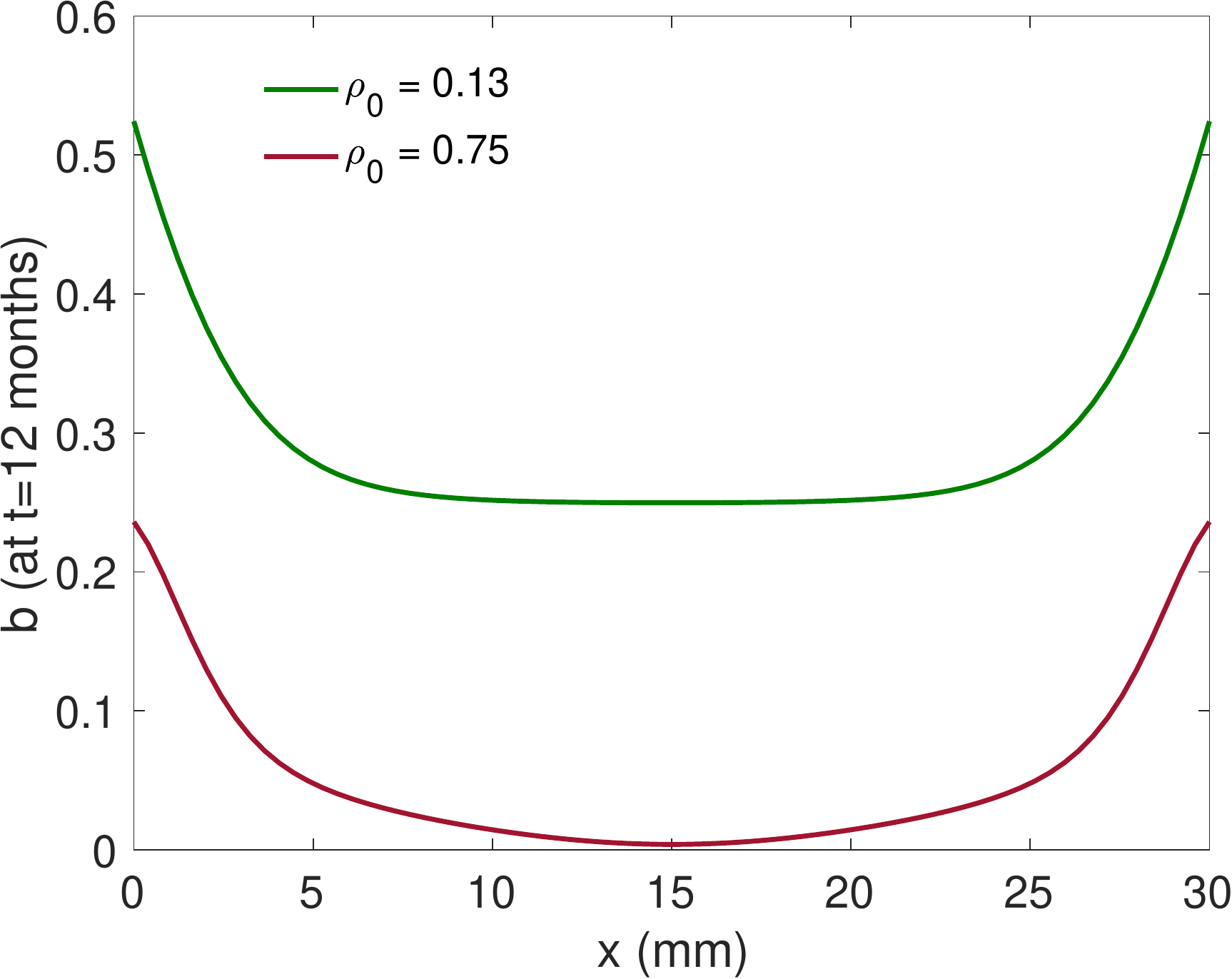}}
\hspace{5mm}
\subfigure[Time evolution of the effective elastic modulus, normalized to the elastic modulus of PCL.]{\label{fig:E_unopt}\includegraphics[width=0.45\textwidth]{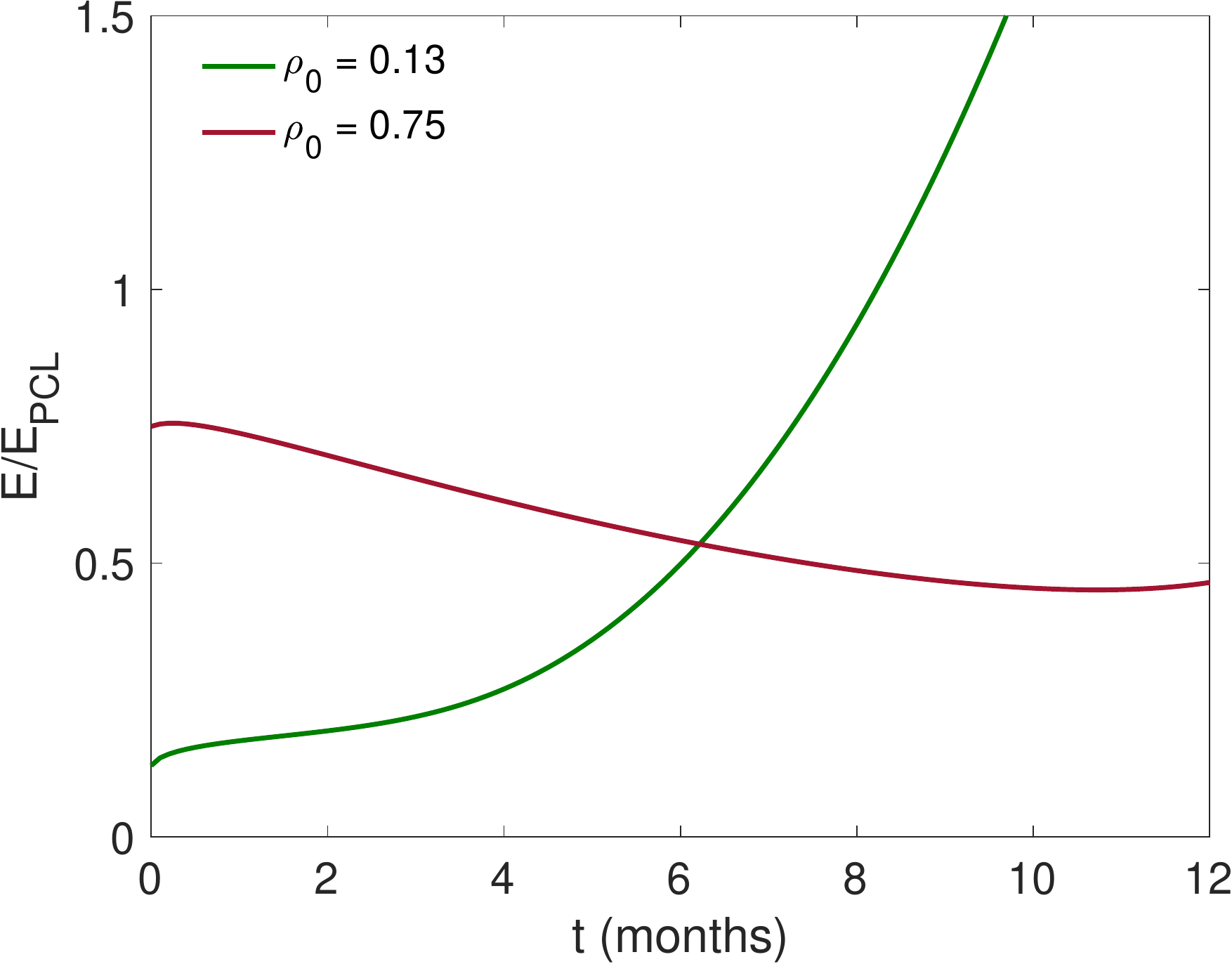}}
\caption{Outcome of the model without optimization.} \label{fig:unopt}
\end{figure}

\begin{table}
\begin{tabular}{l|l}
$T$ & $12$ months \\
\hline
$L$ & $30\mathrm{mm}$ \\
\hline
$\gamma$ & 1\% \\
\hline
$D(\rho)$ & $k_5(1-\rho)$\\
\hline
$\mathbb{C}(\rho,\sigma,b)$ & $\rho\sigma + k_6b$ \\
\hline
$k_1$ & $0.1$ per month \\
\hline
$k_2$ & $80/\gamma$ per month \\
\hline
$k_3$ & $80$ per month \\
\hline
$k_4$ & $0.25/\gamma$ per month  \\
\hline
$k_5$ & $100(\upmu \mathrm{m})^2/\mathrm{s} \approx 260 (\mathrm{mm})^2/\mathrm{month}$ \\
\hline
$k_6$ & $9.0$ \\
\end{tabular}
\caption{Parameters used for the bone regeneration model} \label{tab:params}
\end{table}
The parameters used for the model can be found in Table~\ref{tab:params}. The regeneration time and the length of the scaffold were chosen to model a realistic critical size bone defect, as for example studied in an ovine model in~\cite{Cipitria:2015kp}. The hydration rate constant $k_1$ is set to be consistent with findings in the literature that after one year approximately 30\% of the original molecular weight remains~\cite{Pitt:1981hx}. The diffusivity constant, $k_5$ in $D$, is a standard value for the diffusion of water soluble proteins~\cite{Badugu:2012}, with the $\rho$-dependent prefactor we account for the increased tortuosity of the microscopic diffusion domain due to the scaffold. The constants $k_2$, $k_3$, and $k_4$ are set such that a realistic regeneration outcome is achieved, see for example Figure~\ref{fig:b_unopt} and compare to~\cite[Fig. 2(B), scaffold only]{Cipitria:2015kp}, where the amount of regrown bone in the central region is approximately 75 $\mathrm{mm}^3$, which for a sample diameter of 20 mm and height of 1 mm corresponds to approximately 25\% of regenerated bone matrix for a scaffold density of approximately $\rho_0=13\%$. The relative modulus of bone vs.~scaffold, $k_6$, is chosen to agree with measurements (i.e, the ratio of the nano-indentation hardness for regenerated bone~\cite[Fig. 8(C), scaffold only]{Cipitria:2015kp} of 0.64 GPa and the nano-indentation hardness for PCL~\cite[Table 5]{Lin:2018ey}) of 0.071 GPa. For the elastic modulus $\mathbb{C}$ we use the Voigt Bound for composites.

We note that our model includes local production of bioactive molecules only. An addition of a bone growth booster in the model could be implemented by changing the initial conditions for the function $a$. However, minimizing the required booster does have advantages in particular for patients suffering from bone cancer.

In order to obtain a numerical approximation to the solution of our model, we employ a simple first-order-in-time Euler scheme, where in each time step the discrete equations are solved in the order that they are displayed above (with current values for the quantities substituted in each equation), and are thus decoupled. The diffusion equation for $a$ as well as the elliptic equation for $u$ are discretized in space using piecewise affine finite elements. The resulting ordinary differential equations for $\sigma$ and $a$ are discretized implicitly in time, the equation for $b$ is discretized explicitly in time. The time-step is $0.1$ months and the spatial discretization length is $0.4\mathrm{mm}$.

The resulting regenerated bone matrix for the given parameters and two different constant values for $\rho_0$ are shown in Figure~\ref{fig:b_unopt}. We note that an overly high volume fraction of scaffold material inhibits bone regeneration in our model, which is indeed an observed phenomenon.
 
\section{The optimization algorithm} \label{sec:opt}

Using current additive manufacturing technologies, it is possible to generate  unit-cell based scaffolds with any given porosity distribution $\rho_0$ on the domain occupied by the implant. We note that this distribution of porosity is a major aspect of the scaffold architecture. The solution of the evolution equations in our model now allows the deduction of, for example, the rate of proliferation of newly generated bone matrix within the implant, all dependent on the initial scaffold porosity distribution.

The model above can furthermore be used to evaluate the time evolution of the mechanical stiffness (i.e., effective elastic modulus of the structure stemming from both the scaffold material and the regenerated bone) of the implant for a given $\rho_0$. This mechanical stiffness is (in our case) proportional to the total elastic energy in the system at time $t\in [0,T]$,
\[
E^\text{el}(t) = \inf_{\stackrel{v\in H^1((0,L))}{v(0)=0, v(L)=\gamma L}} \int_0^L v_x\mathbb{C}(\rho(x,t),\sigma(x,t),b(x,t))v_x \,\mathrm{d}x,
\]
where $\rho$, $\sigma$, and $b$ follow the equations above (and thus depend on $\rho_0$). Figure~\ref{fig:E_unopt} shows the evolution of $E$ in time for the two unoptimized test cases.

It is possible to maximize the minimum over the regeneration time of the overall mechanical stiffness
\[
E^\text{min}(\rho_0) = \min_{t\in [0,T]} E^\text{el}(t)
\]
 of the scaffold among all physiologically suitable (i.e., with a cutoff to ensure $\rho_0(x)\le 0.7$ in order to not prevent vascularization) initial scaffold porosity distributions.
 
 It is important to note that for PCL scaffolds such an optimization is a reasonable approach. For certain titanium-mesh scaffolds it has been found that softer scaffolds yield better regeneration results~\cite{Pobloth:2018}---PCL, however, is already a material that is less stiff than bone. Therefore, it is not an issue here to obtain the mechanical stimulus that is required for bone regeneration. Instead, the main drawback of PCL scaffolds is their potential for mechanical failure, which can be minimized by maximizing their mechanical stiffness.
 
 We achieve this with the established method of using a gradient flow type optimization of the objective function $E^\text{min}$. To be more precise, we calculate an approximation of the Fr{\'e}chet derivative $\delta_{\rho_0}E^\text{min}(\rho_0)$ by using finite differences in each node of the spatial discretization. Then a gradient ascent with a soft pointwise constraint on the volume fraction is performed. The computational cost for such an optimization method is easily manageable due to the use of coarse grained values in the evolution equation system. Our MATLAB implementation performs a full optimization and reaches a stationary state of the gradient ascent in a few minutes on a regular desktop computer, thus more sophisticated methods were not deemed necesseary. Our method can therefore yield an optimized scaffold design, adapted to a given mechanical load and other (possibly patient specific) parameters.
 
\section{Numerical Results} \label{sec:num}

\begin{figure}
\centering 
\subfigure[Volume fraction distributions.]{\label{fig:rho_all}\includegraphics[width=0.45\textwidth]{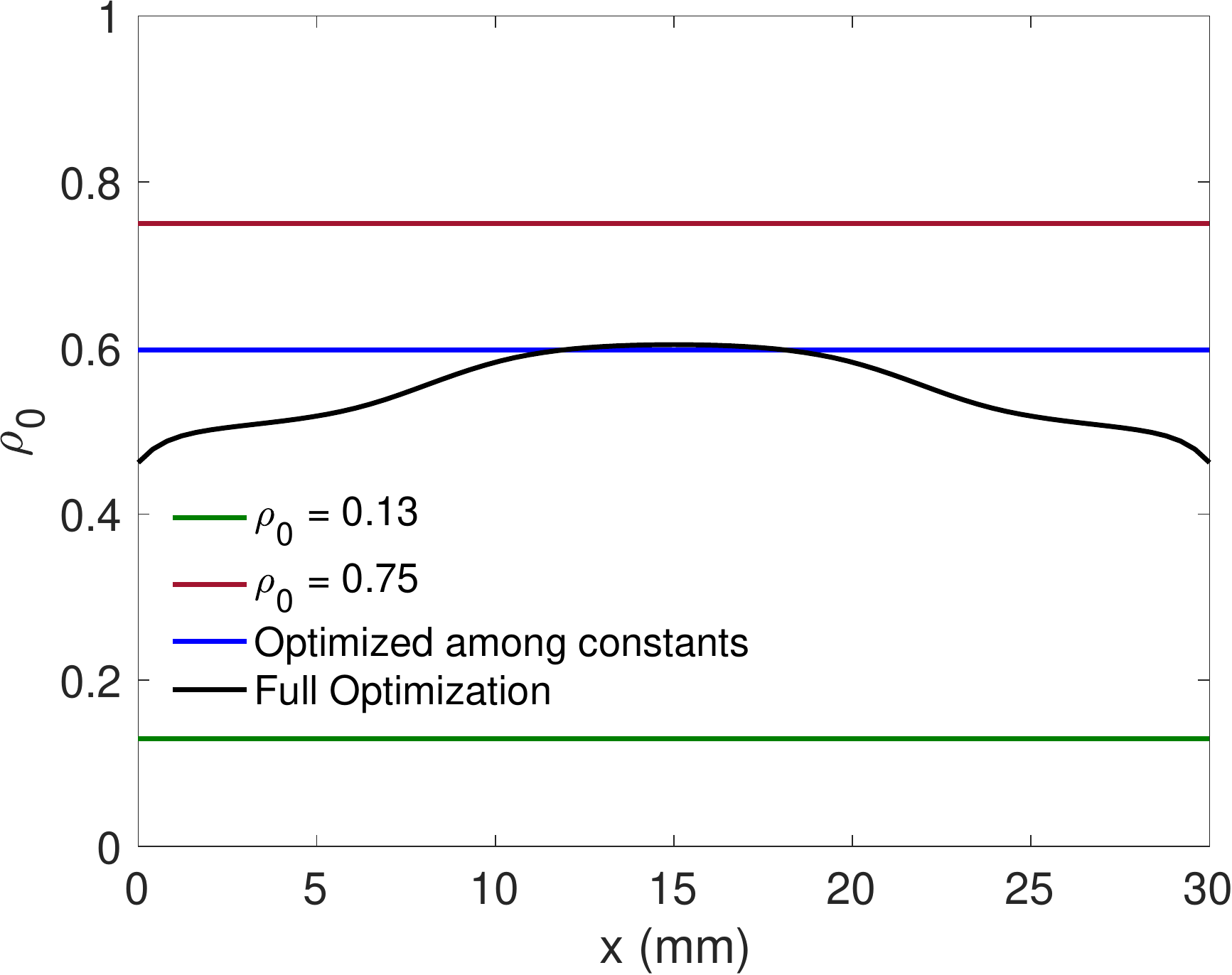}}
\hspace{5mm}
\subfigure[Time evolution of the normalized effective elastic modulus depending on $\rho_0$.]{\label{fig:e_all}\includegraphics[width=0.45\textwidth]{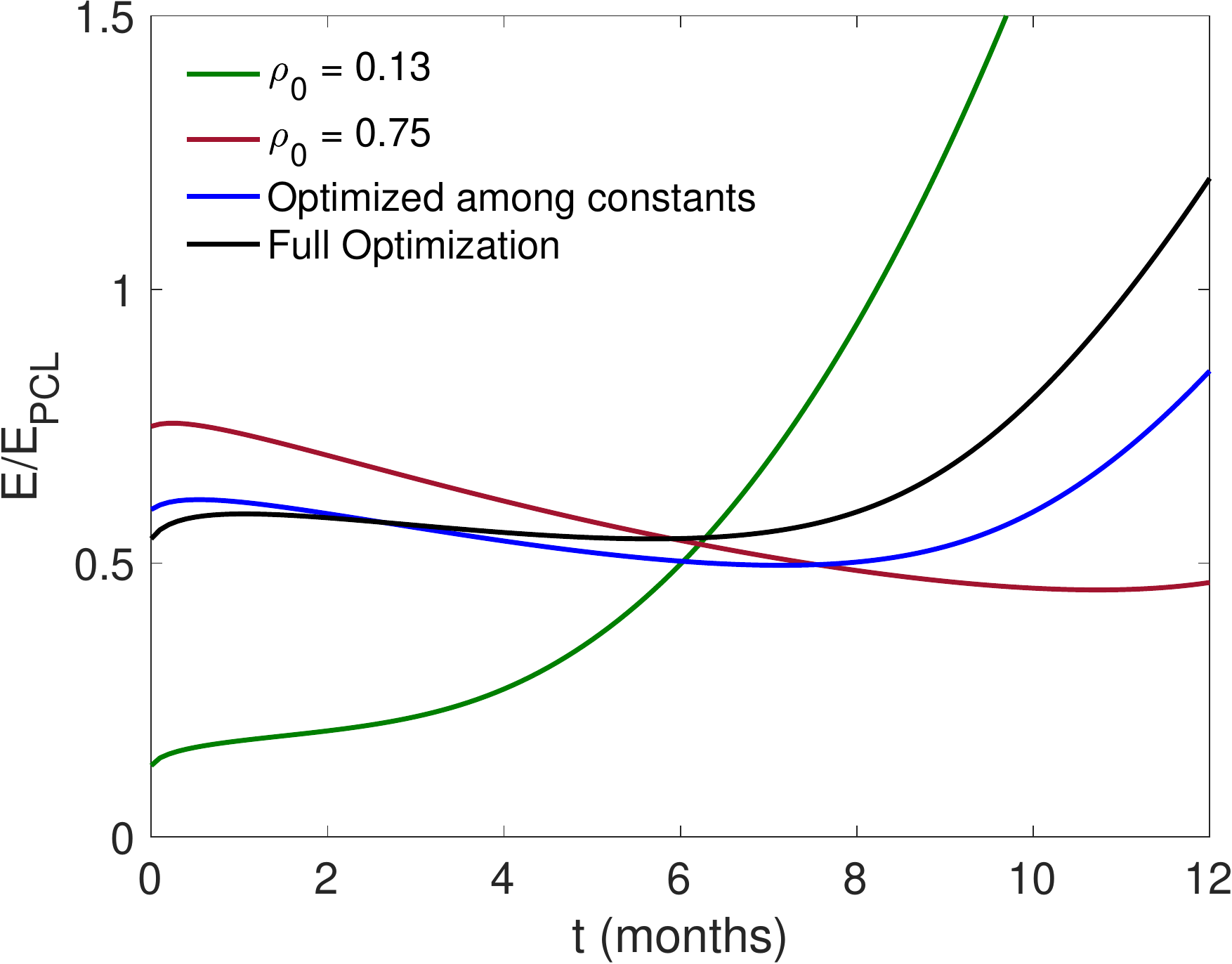}}
\caption{Comparison of scaffold optimization results.} \label{fig:optimized}
\end{figure}

\begin{table}
\begin{tabular}{l|l}
$\rho_0 = 0.13$ & 0.13 \\
\hline
$\rho_0 = 0.75$ & 0.45 \\
\hline
Optimization among constant $\rho_0$ & 0.50 \\
\hline
Full optimization & 0.54 \\
\hline
Optimization among constant $\rho_0$ (reduced bone regeneration) &  0.26 \\
\hline
Full optimization (reduced bone regeneration) &  0.36 \\ 
\hline
Reduced bone regeneration but $\rho_0$ optimized for regular patient  & 0.28 \\
\end{tabular}
\caption{Comparison of minimal elastic modulus $E^\text{min}(\rho_0)$ for the scaffold designs under consideration. The values are normalized such that the scaffold material PCL has an elastic modulus of 1.} \label{tab:opt_values}
\end{table}

Figure~\ref{fig:optimized} shows a comparison of the optimized scaffold designs. Table~\ref{tab:opt_values} shows the minimal elastic modulus values for all designs. In particular, we also compare the optimization among constant-volume fraction scaffolds. In our model, with the parameters given in Table~\ref{tab:params}, the difference in the minimal elastic modulus between the full optimization and the optimization restricted to constant volume fractions is approximately 8\%. In particular, we see that the optimal scaffold is denser in the middle region---this can of course be understood from heuristic arguments: the regenerated bone grows back from the ends where the scaffold is attached to the intact bone matrix. Thus, in the central region, the scaffold polymer has to maintain the structural integrity for a longer time by itself, while undergoing bulk erosion.

\begin{figure}
\centering 
\subfigure[Optimized volume fraction distributions.]{\label{fig:rho_all_osteo}\includegraphics[width=0.45\textwidth]{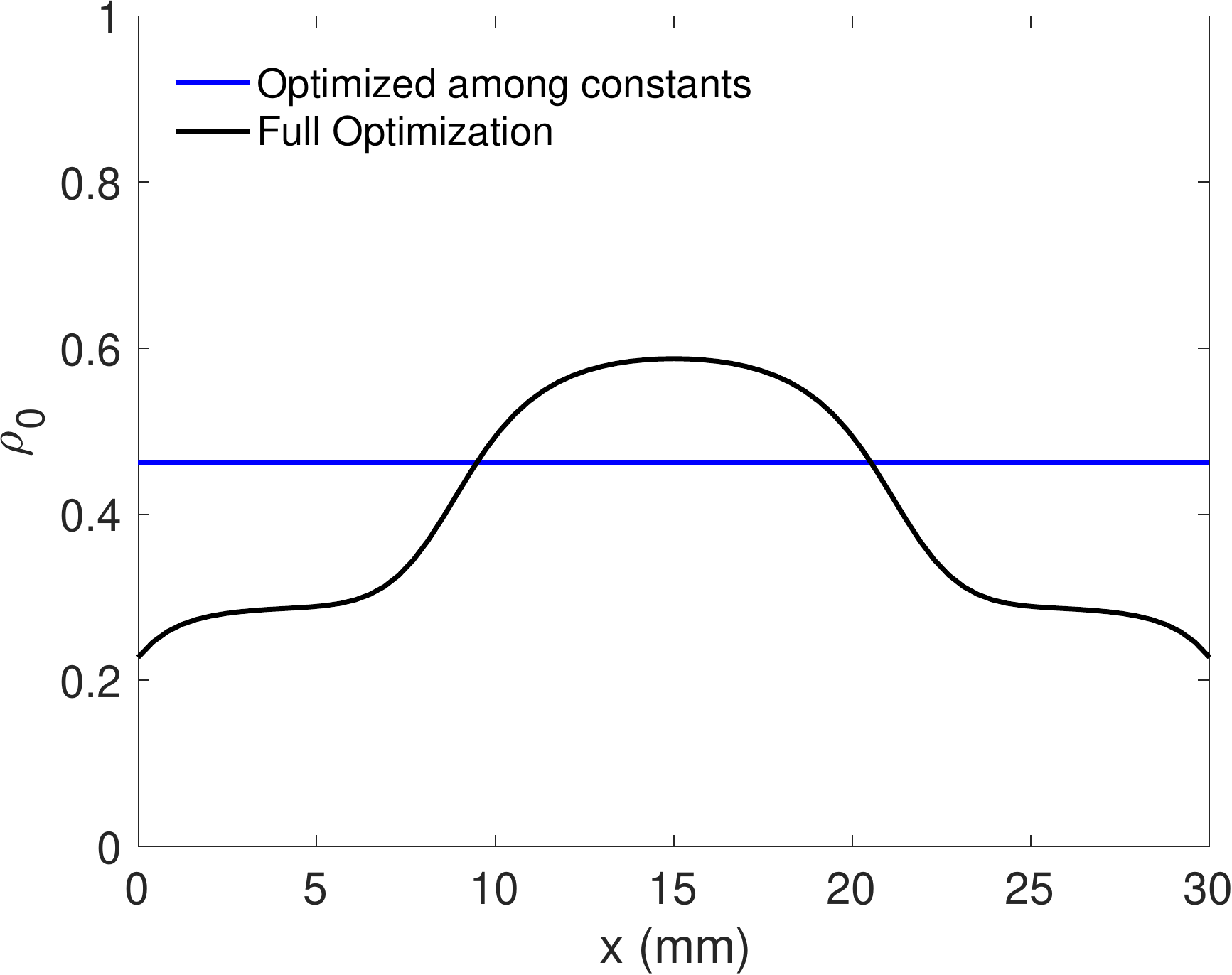}}
\hspace{5mm}
\subfigure[Time evolution of the normalized effective elastic modulus depending on $\rho_0$.]{\label{fig:e_all_osteo}\includegraphics[width=0.45\textwidth]{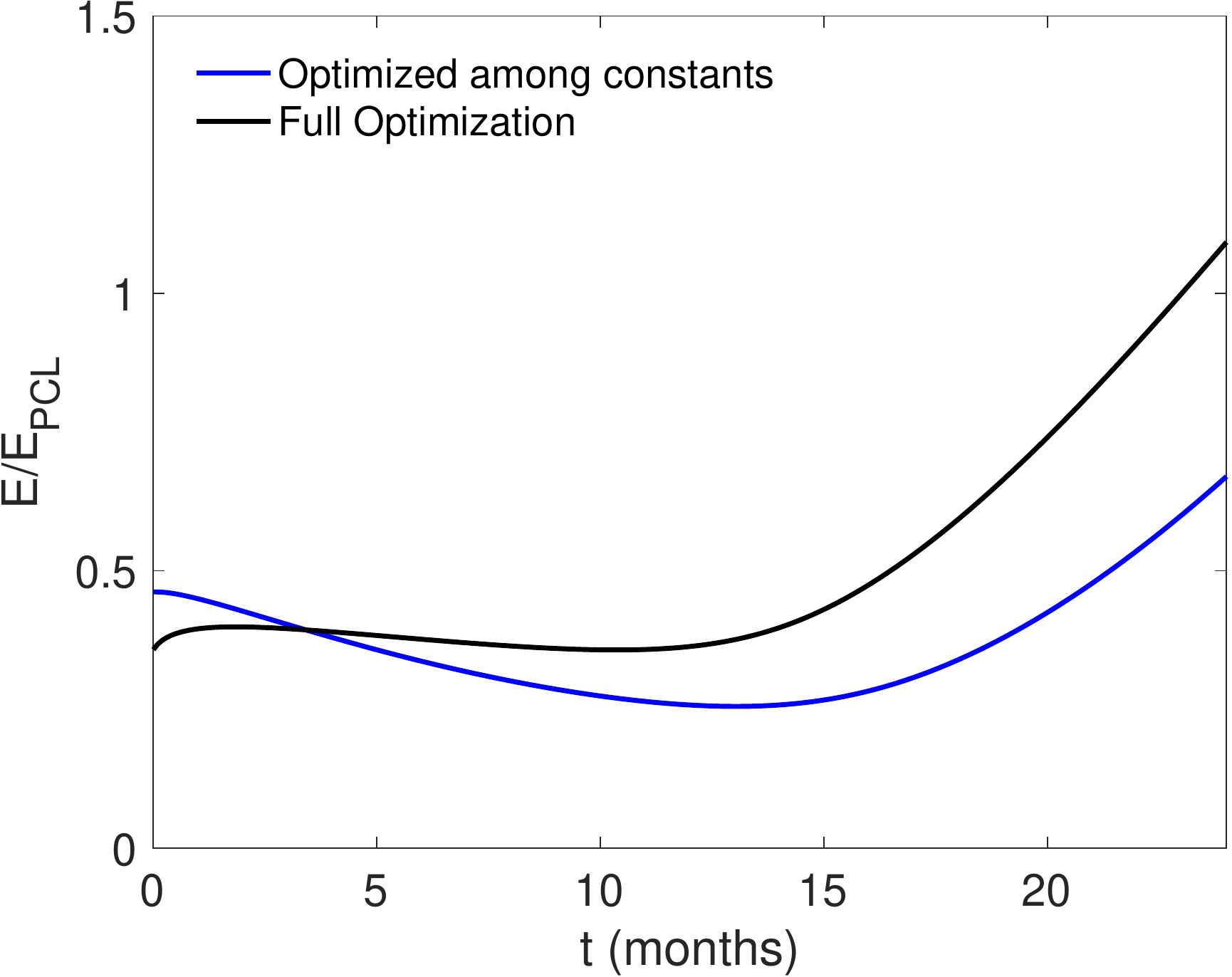}}
\caption{Comparison of scaffold optimization results for the model with reduced bone regeneration, i.e., $k_4=0.125/\gamma$ and  $k_6=4.5$.} \label{fig:osteo}
\end{figure}

Furthermore, we conducted a numerical experiment where the bone-re\-generation coefficient was reduced to $k_4=0.125/\gamma$ and the stiffness ratio $k_6=4.5$, to emulate a patient with reduced regeneration capacity, for example due to osteoporosis. Table~\ref{tab:opt_values} shows that the difference between the full optimization and the optimization among constant $\rho_0$ grows to nearly 40\% in this case. We also note that the scaffold optimized for the `regular' value of $k_4$ and $k_6$ is far from optimal for reduced bone regeneration. The optimized $\rho_0$ and corresponding elastic moduli are plotted in Figure~\ref{fig:osteo}.

\section{Conclusions} \label{sec:conc}
We have devised a simple model for bone regeneration that is suitable for use in an optimization routine for polymer bone scaffolds. As a proxy for the mechanical stability of such scaffolds, we maximize the effective elastic modulus of the combination of scaffold- and regenerated bone material, after taking its minimum over the regeneration time. With this method, it is possible to find optimized volume fraction distributions for additively manufactured scaffolds based, e.g., on periodic unit cell designs. Within our model, compared to an optimization among only constant volume fractions, a small increase of 8\% for the objective function can be achieved by optimizing among non-constant initial volume fraction distributions.

We remark that, of course, our model is very simplistic and a number of parameters are not known very exactly. However, the general outcome of optimal scaffold designs that increase their volume fraction in their central region (in order to provide more stability until the regenerated bone reaches the center) and decrease their volume fraction at the edges (in order to not impede bone-ingrowth) holds for a large range of different parameters.

In a test case with reduced bone regeneration ability, the difference is significantly increased. This shows that our model can benefit tremendously from patient-specific parameters to produce patient-specific optimal scaffold designs.

Presently, diagnosis of bone defects is highly dependent on X-ray micrographs, which only provide visual guidance for treatment and management but lack precise information on the patient's intrinsic bone regeneration capability. Potentially, this could be overcome by leveraging on Omics technology and multi-model analysis through bioinformatics techniques to improve patient stratification in term of intrinsic bone regeneration capability while unraveling the underlying biological mechanisms that govern the bone regeneration cascade. The inclusion of such patient-specific parameters would significantly improve the herein proposed optimization model for personalized bone scaffold designs. Nonetheless, such an endeavor will require long-term synergistic and complementary research efforts across multitude of disciplines. 

Hence, within the scope of design optimization model, future work includes a refinement of the model in order to be better aligned with experimental results as well as an extension to a full three-dimensional optimization.
\bibliographystyle{alphaabbr}
\bibliography{references}

\end{document}